\documentclass[aps,pre,showpacs]{revtex4}
\usepackage{graphicx}
\usepackage{colordvi}
\begin{document}

\title{Matter wave soliton bouncer}

\author{A. Benseghir$^1$, W. A. T. Wan Abdullah$^1$, B. B. Baizakov$^2$, and F. Kh. Abdullaev$^3$}
\affiliation{
$^1$ Department of Physics, Faculty of Science, University of Malaya, 50603, Kuala Lumpur, Malaysia \\
$^2$ Physical-Technical Institute, Uzbek Academy of Sciences, 100084, Tashkent, Uzbekistan \\
$^3$ Department of Physics, Faculty of Science, International
Islamic University of Malaysia, 25200, Kuantan, Malaysia }
\date{\today}

\begin{abstract}
Dynamics of a matter wave soliton bouncing on the reflecting
surface (atomic mirror) under the effect of gravity has been
studied by analytical and numerical means. The analytical
description is based on the variational approach. Resonant
oscillations of the soliton's center of mass and width, induced by
appropriate modulation of the atomic scattering length and the
slope of the linear potential are analyzed. In numerical
experiments we observe the Fermi type acceleration of the soliton
when the vertical position of the reflecting surface is
periodically varied in time. Analytical predictions are compared
with the results of numerical simulations of the Gross-Pitaevskii
equation and qualitative agreement between them is found.
\end{abstract}
 \maketitle

\section{Introduction}

A particle bouncing on the reflecting surface under the effect of
gravity represents one of the analytically solvable models in
quantum mechanics \cite{sakurai,flugge}. Gibbs introduced the name
``quantum bouncer" \cite{gibbs} for the object, and it was
extensively studied in many articles of pedagogical orientation
\cite{gea-banacloche1999,langhoff} and original research papers
(for a recent review see \cite{belloni2014}). The practical
interest in this model has emerged from recent experiments aimed
at probing the coherence properties of Bose-Einstein condensates
falling under gravity and bouncing off a mirror formed by a
far-detuned sheet of light \cite{bongs}, quantum reflection of
matter waves \cite{pasquini}, and measuring the Casimir-Polder
force acting upon the atoms near solid surfaces \cite{harber2005}.

Another important result linked to the quantum bouncer problem has
been the experimental observation of quantum bound states of
neutrons in the Earth's gravitational field
\cite{nesvizhevsky,abele2012, ichikawa2014}. In these pioneering
experiments the quantum states of matter formed by a gravitational
field were observed for the first time. Also, the model is of
particular interest from the viewpoints of the physics and
applications of quantum states of nanoparticles in the vicinity of
surfaces \cite{durand2011}. An optical analogue of the quantum
bouncer, a photon bouncing ball, was experimentally demonstrated
using the circularly curved optical waveguide \cite{della-valle}.
Significance of the model for the study of the dynamics of
particles in quantum-classical interface was pointed out in
\cite{robinett}.

In this work we extend the quantum bouncer model to the nonlinear
domain by considering the dynamics of a matter wave soliton
governed by the Gross-Pitaevskii equation (GPE). The linear
potential entering the GPE represents the Earth's gravitational
field acting on the soliton in vertical direction, while the
horizontal atomic mirror \cite{atom_mirror} created by a laser
beam or magnetic field stands for the reflecting surface. The
matter wave soliton performs bounded motion in such a
gravitational cavity. The effect of nonlinearity, originating from
the atomic interactions in Bose-Einstein condensate (BEC), shows
up as an ability of the bouncing wave packet to remain localized
during the evolution, behaving like a rigid ball, rather than a
deformable wave packet. The possibility of tuning the atomic
interactions in the condensate by external magnetic
\cite{kohler2006} and optical \cite{ciurylo2005} fields opens
perspectives in exploring the bouncer problem in both the quantum
and classical limits.

Our main objective is to develop analytical description of the
soliton's dynamics above the atomic mirror under the effect of
gravity. As an illustration of the developed model we consider the
resonant oscillations of the soliton's center of mass position
under periodically varying strength of nonlinearity and the slope
of the quasi-1D trap with respect to the horizontal reflecting
surface. The strength of nonlinearity can be tuned using the
Feshbach resonance technique \cite{kohler2006}, or alternatively,
by changing the strength of the radial confinement. In numerical
simulations we demonstrate the Fermi type acceleration of the
soliton when the vertical position of the mirror is periodically
varied in time. It should be noted that Fermi acceleration of
matter wave packets was previously considered in \cite{saif} for
the case of non interacting BEC, in the setting where matter wave
solitons do not exist. In these works non-dispersive acceleration
of the wave packet was reported to take place under certain
conditions, when the modulation strength and frequency provide the
dynamical localization of the matter wave.

The advantage of the present setting is that, the bouncing matter
wave packet preserves its integrity due to the focusing
nonlinearity of the BEC, which counteracts the dispersive
spreading. Another interesting approach to acceleration of a
single quantum particle, also feasible in the context of matter
waves, was reported in \cite{granot2011}. The mechanism consists
in binding the wave packet by a delta function potential well and
involving in accelerated motion along with the potential. In the
linear case and ideal mirror potential our model reduces to the
equation which has analytic solution in terms of Airy functions.
The dynamics of Airy beams currently represents one of the
actively explored topics motivated by important applications in
optical communications and nonlinear optics \cite{airy}.

The paper is structured as follows. In Sec. II we introduce the
mathematical model and illustrate the distinctive features of the
nonlinear model as compared to its linear counterpart. In Sec. III
a variational approach for analytical treatment of the nonlinear
model has been developed and its predictions are compared with
numerical simulations of the original GPE. Sec. IV is devoted to
exploring the resonant oscillations of the wave packet above the
mirror, and Fermi type of acceleration of matter wave solitons
when the vertical position of the reflecting surface is
periodically varied in time. In Sec. V we summarize our findings.

\section{The model and main equations}

The Bose-Einstein condensate is a giant matter wave packet which
is strongly affected by gravity. In particular, a matter wave
packet released from the trap falls towards Earth like a bunch of
coherent atoms. The effect of gravity is essential for the
operation of atom lasers \cite{mewes1997}.

In the present model the gravitational field acting on atoms in
the vertical direction and a horizontal atom mirror which reflects
them back, form a cavity for the matter wave packet. Below we
consider the motion of a matter wave soliton within such a
gravitational cavity. The model is based on the following one
dimensional GPE
\begin{equation}\label{gpe1}
i\hbar \frac{\partial \psi}{\partial t}=-\frac{\hbar^2}{2
m}\frac{\partial^2 \psi}{\partial x^2} + m g x \psi + U(x) \psi +
2\hbar \omega_{\bot} a_s |\psi|^2 \psi,
\end{equation}
where $\psi(x,t)$ is the wave function of the condensate trapped
in a tight quasi-1D trap, $x$ is the spatial coordinate of the
wave packet above the horizontal atomic mirror, represented by the
reflecting potential $U(x)$, $g$ is the strength of the
gravitational potential, $\omega_{\bot}$ is the trap frequency in
the tightly confining radial direction, $m, a_s$ are the atomic
mass and $s$ - wave scattering length, respectively.

The gravitational units of space and time, defined as
\begin{equation}
l_g = \left(\frac{\hbar^2}{m^2 g}\right)^{1/3}, \qquad t_g =
\left(\frac{\hbar}{m g^2}\right)^{1/3},
\end{equation}
allow to rewrite the Eq. (\ref{gpe1}) in the dimensionless form
\begin{equation}\label{gpe2}
i\psi_t + \frac{1}{2} \psi_{xx} + \gamma |\psi|^2 \psi - \alpha \,
x \psi + V(x) \psi = 0,
\end{equation}
where the new variables are defined as $x \rightarrow x/l_g$, \ $t
\rightarrow t/t_g$, \ $V(x) = - U(x)/(m g l_g)$, $\psi \rightarrow
\sqrt{2 \omega_{\bot} |a_s| t_g} \psi$. Here we took into regard
that for BEC with attractive atomic interactions, $a_s < 0$. In
Eq. (\ref{gpe2}) the linear potential term ($\sim x$) accounts for
the effect of gravity, while the atomic mirror is represented by
$V(x)$. We introduced an additional parameter $\alpha =
\sin(\beta)$ to account for the possibility of altering the effect
of gravity by changing the angle $\beta$ formed by the axis of the
quasi-1D waveguide and the horizontal reflecting surface. For
vertical position ($\beta=\pi/2$) of the waveguide $\alpha = 1$,
at smaller angles $ 0 < \beta < \pi/2 $, then $0 < \alpha < 1$.
Such a setting is of interest in view of recent research on the
behavior of BEC in microgravity \cite{microgravity} and the
quantum reflection of matter waves \cite{pasquini}, where the cold
atoms should approach the attractive potential at very low
velocity. Similarly, the additional parameter $\gamma$ can be used
for nonlinearity management $\gamma(t) = a_s(t)/a_s^0$, then in
the normalization for $\psi$ in Eq. (\ref{gpe2}) the background
value of $a_s^0$ should be assumed. The following two cases will
be relevant to our further analysis
\begin{equation}\label{mirror}
\mbox{(a) ideal mirror} \qquad V(x)=\left\{
\begin{tabular}{cc}
      0,   & if $x \geq 0$ \\
$+\infty$, & if $x < 0$
\end{tabular}
\right., \qquad \mbox{(b) weakly transparent reflecting surface}
\qquad V(x) = V_0 \delta(x),
\end{equation}
where $\delta(x)$ is the Dirac delta function which has been
multiplied by the strength $V_0$.

A detailed study of the wave packet dynamics described by Eq.
(\ref{gpe2}) in the linear model ($\gamma=0$) for ideal mirror was
reported in Ref. \cite{gea-banacloche1999}. Before proceeding to
analytical description of the nonlinear model ($\gamma = 1 $) it
is instructive to compare these two limits by numerical
simulations of the governing equation (\ref{gpe2}). Such a
preliminary study will help to elucidate the effect of
nonlinearity on the dynamics of a wave packet bouncing above the
atomic mirror.

\begin{figure}[htb]
\centerline{
\includegraphics[width=8cm,height=6cm,clip]{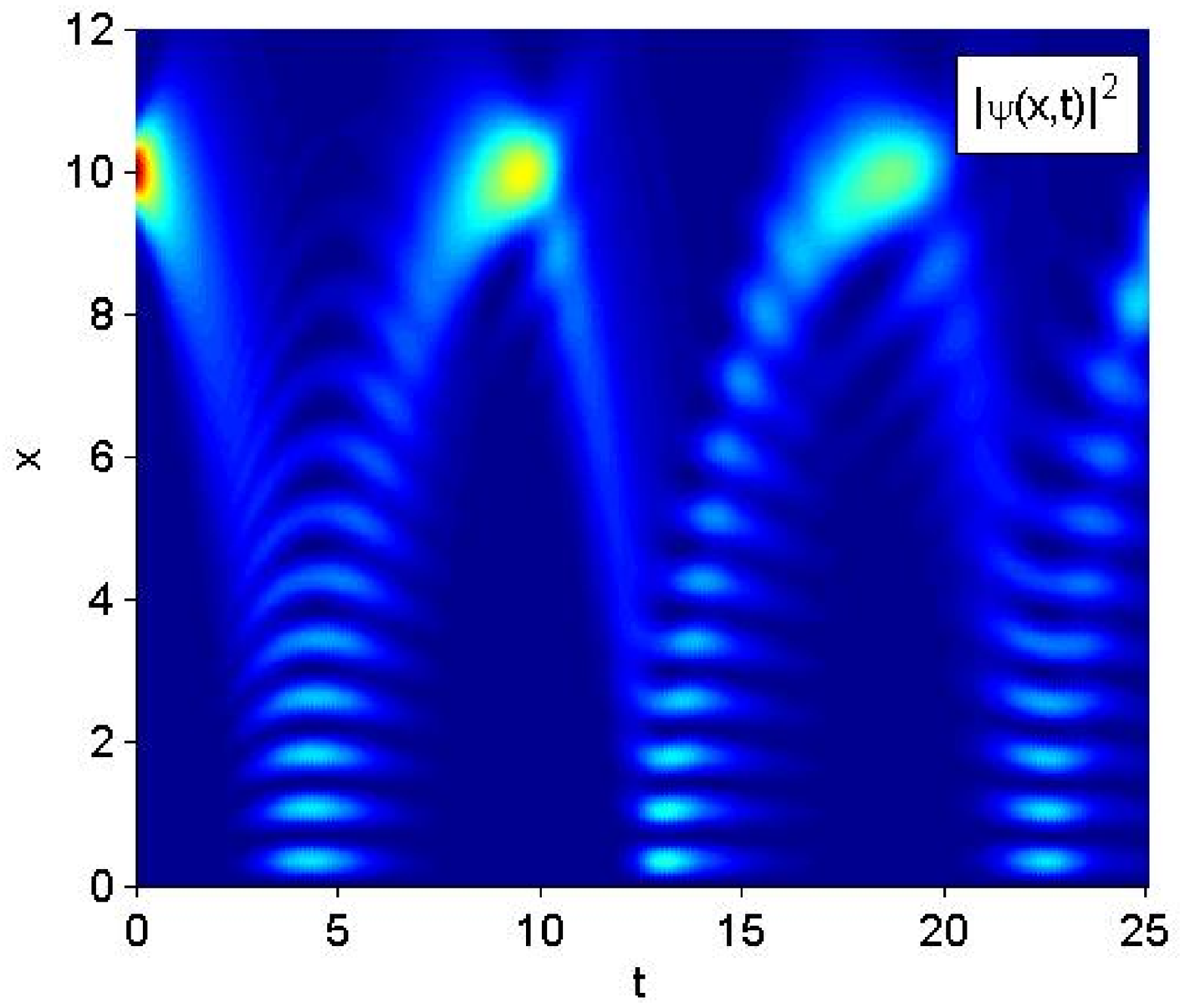} \qquad
\includegraphics[width=8cm,height=6cm,clip]{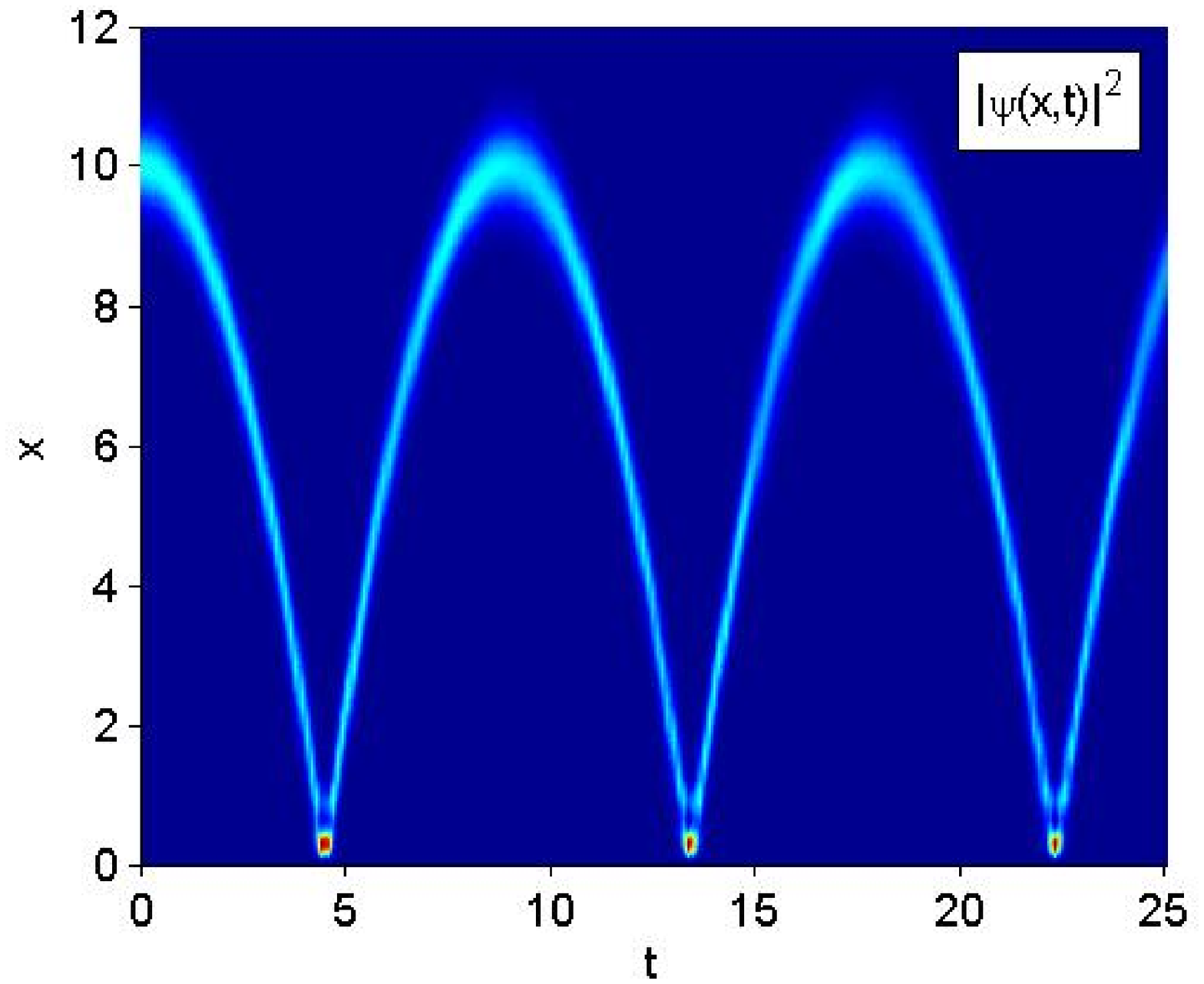}}
\caption{(Color online) Left panel: The first three bouncing of
the wave packet from the ideal mirror for the linear model
($\gamma = 0$) is shown through the density plot $|\psi(x,t)|^2$.
Right panel: The same for the nonlinear model ($\gamma = 1$). In
both cases a wave packet $\psi(x,0) = A \exp(-(x-x_0)^2/a^2)$ with
$A=2, \ a=0.8$ and $x_0 = 10$ has been employed as initial
condition for the governing~Eq.(\ref{gpe2}).} \label{fig1}
\end{figure}

\begin{figure}[htb]
\centerline{
\includegraphics[width=3.5cm,height=6cm,clip]{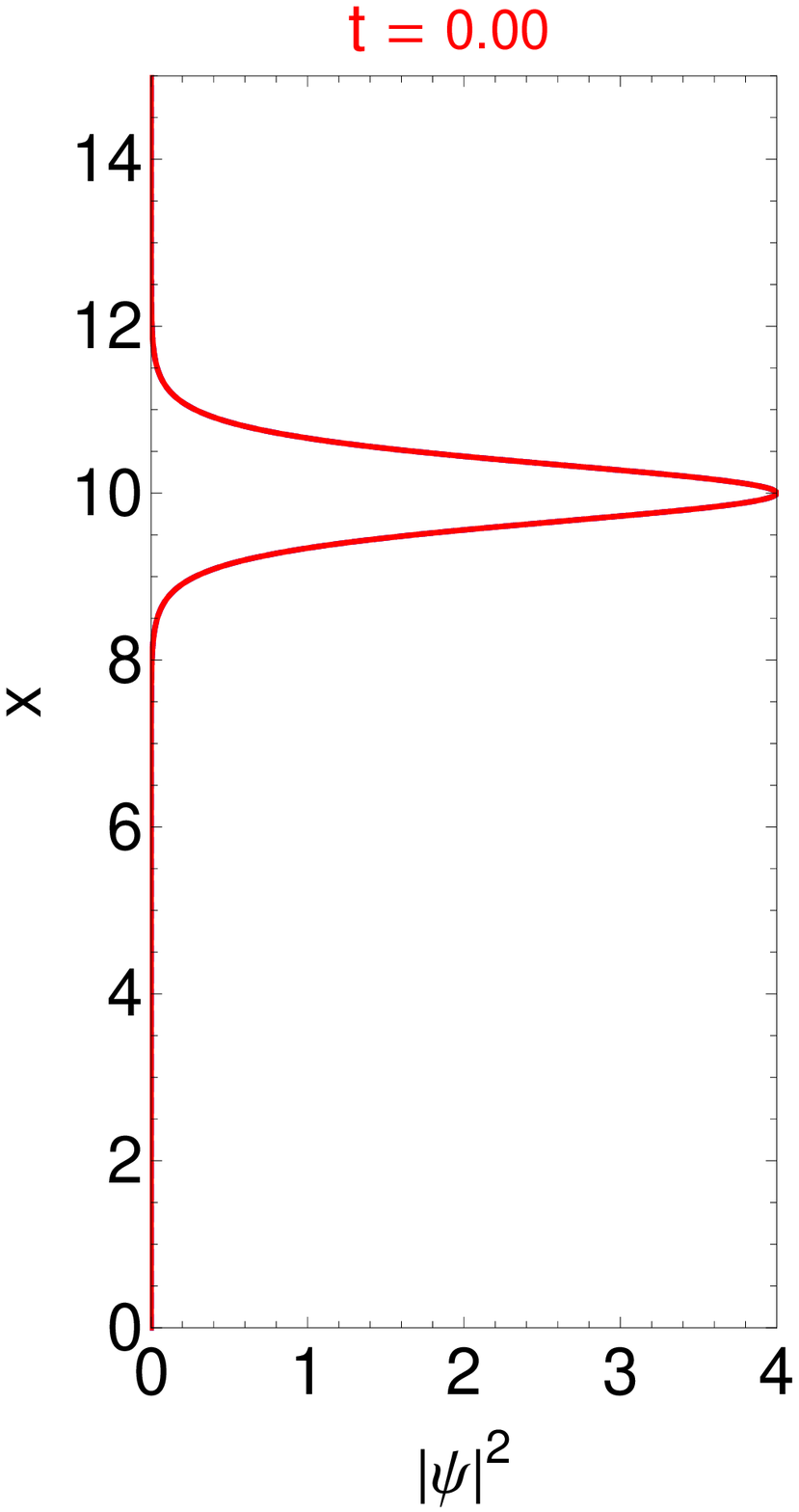}
\includegraphics[width=3.5cm,height=6cm,clip]{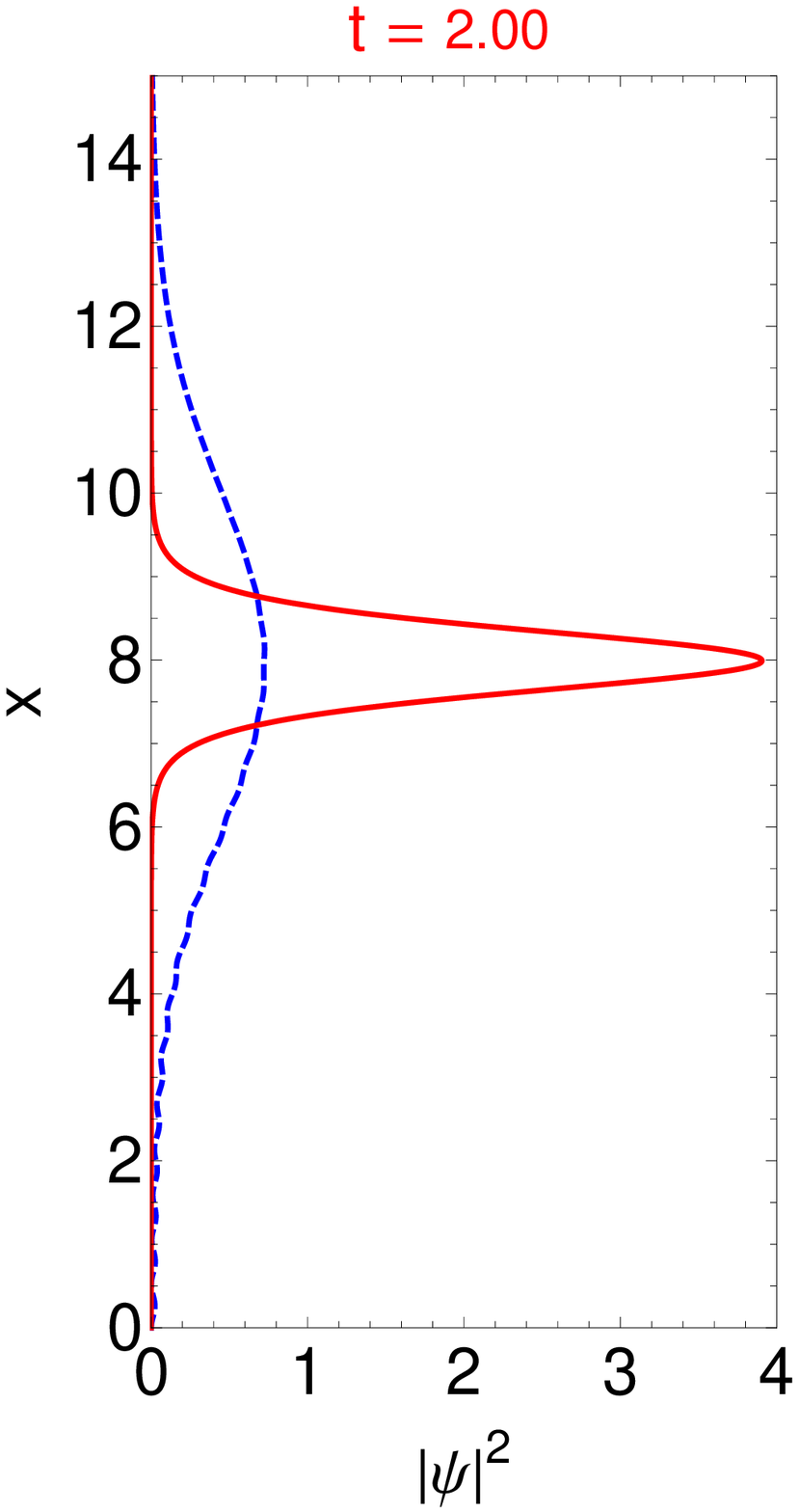}
\includegraphics[width=3.5cm,height=6cm,clip]{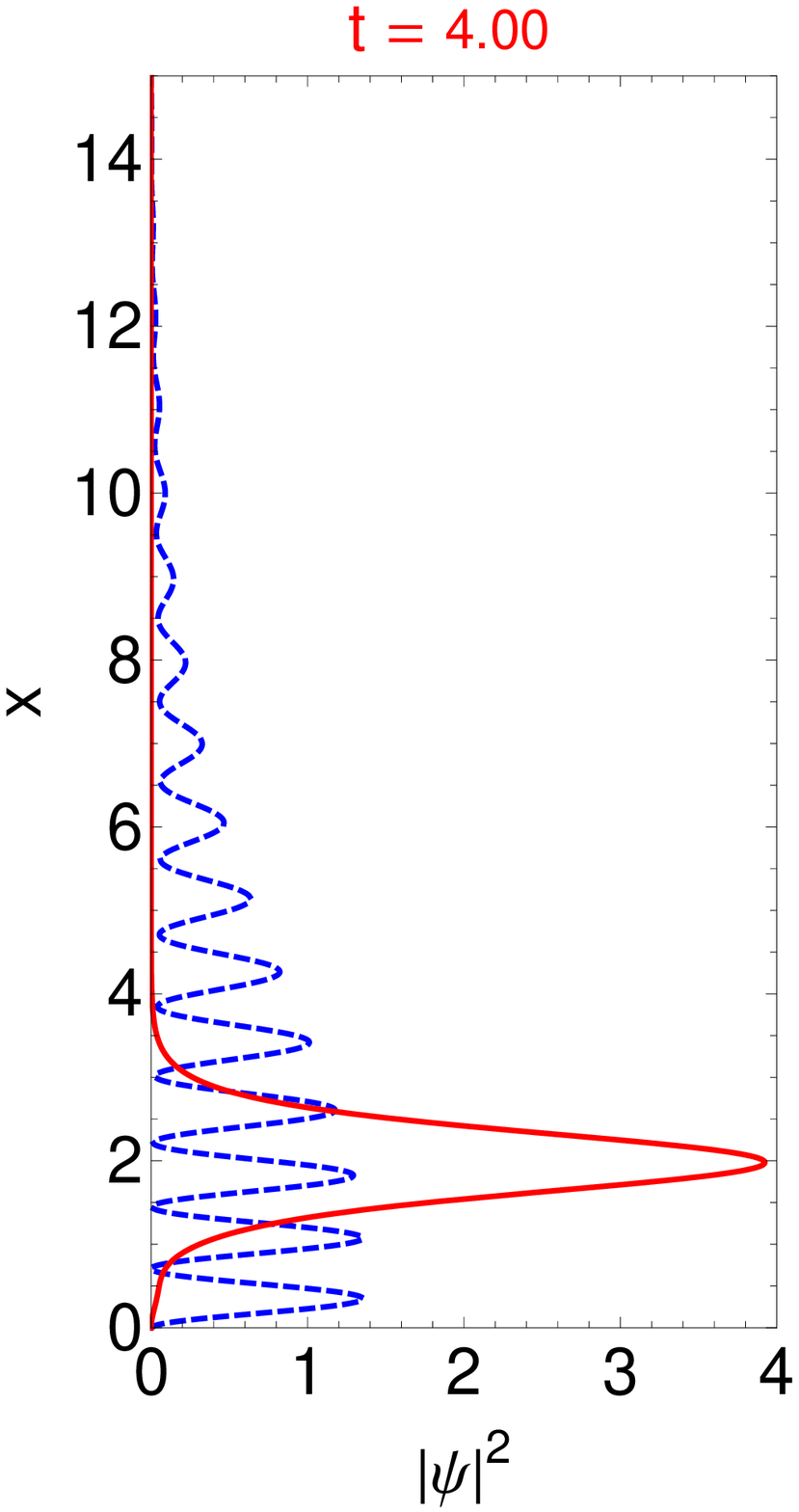}
\includegraphics[width=3.5cm,height=6cm,clip]{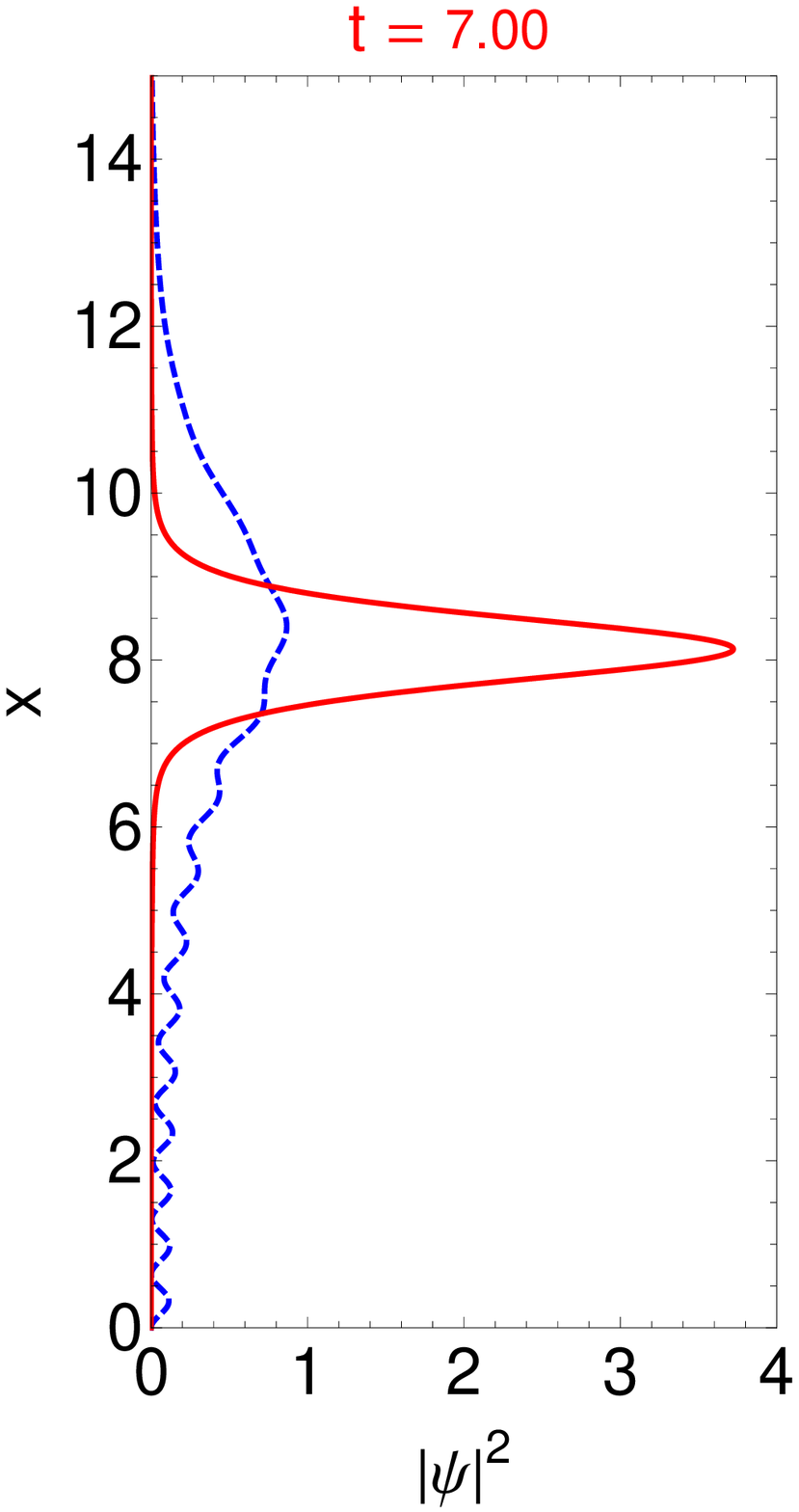}
\includegraphics[width=3.5cm,height=6cm,clip]{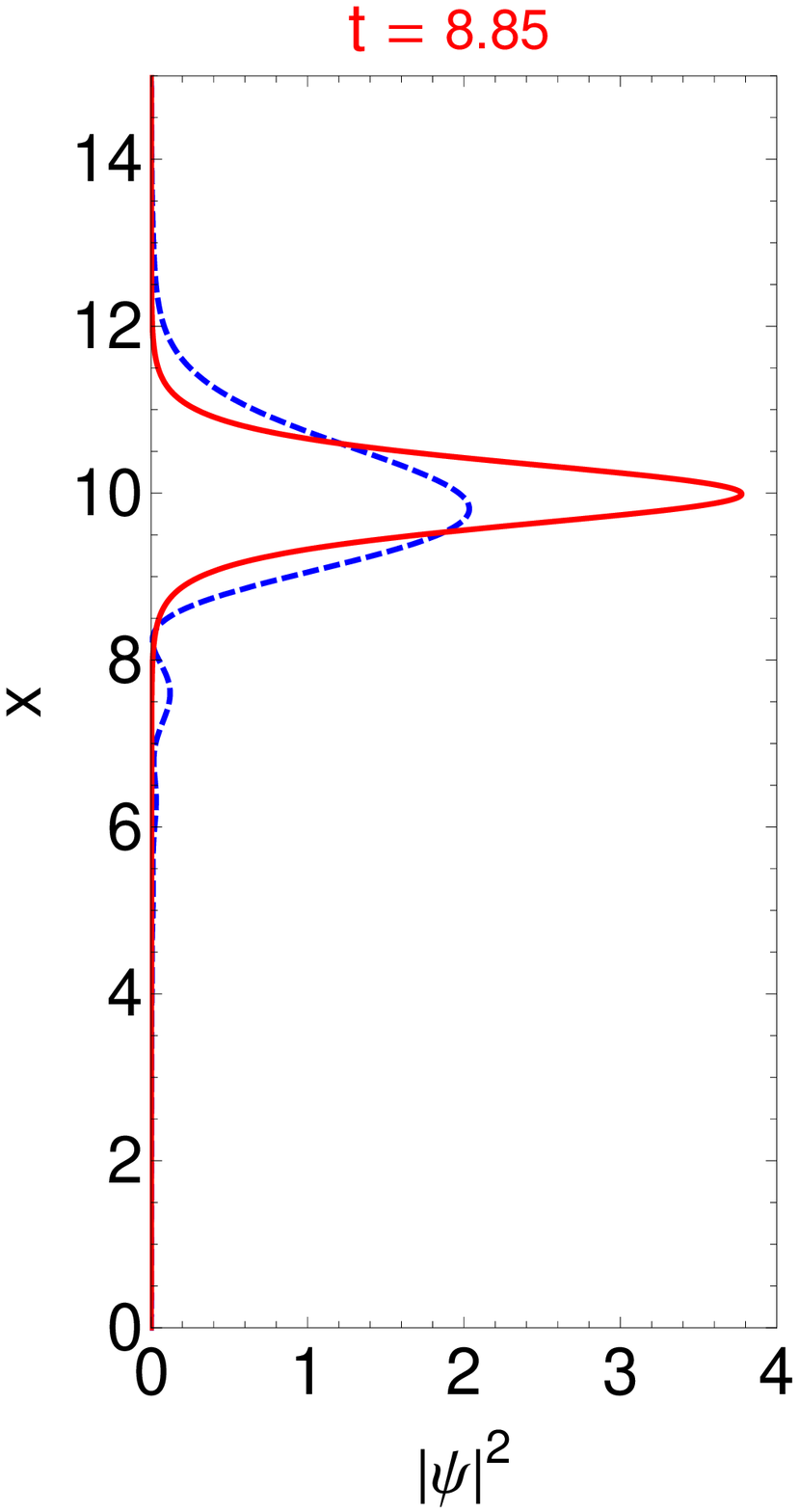}}
\caption{(Color online) Snapshots of the wave packet at different
times (shown on top of each figure) during one bouncing period. In
the linear model the wave packet quickly expands and shows strong
interference with waves reflected by the mirror placed at $x=0$
(blue dashed line). At final time $T_b$ the wave packet does not
fully recover its initial form. In the nonlinear case the wave
packet keeps its integrity during the evolution period and almost
fully recovers its initial form at $T_b$ (red solid line). All
parameters are the same as in the previous figure.} \label{fig2}
\end{figure}

In Fig. \ref{fig1} we illustrate the features of the linear and
nonlinear models for the dynamics of the wave packet dropped from
the height $x_0=10$ above the mirror positioned at $x=0$. The main
difference appears to be enhanced spreading of the wave packet and
strong interference with reflected waves in the linear model, as
compared to the nonlinear case, where these phenomena do not show
up.
The distinctions between the two models is clearly observed in
Fig. \ref{fig2}, where we compare the corresponding wave profiles
at different times during one period of bouncing $T_b$, which is
estimated from classical equation $d^2 x/d t^2 = -g$. In
dimensionless units introduced for Eq. (\ref{gpe2}) we need to set
$g=1$. Then a classical particle dropped from the height $x_0$
reaches the ground at $t_b = \sqrt{2 x_0}$, therefore the
classical bouncing period is $T_b = 2 t_b = 2 \sqrt{2 x_0}$.

To estimate the parameters of the model we consider the $^{85}$Rb
condensate, for which $a_s=-20$ nm, $l_g \approx 1.3 \,\mu$m, $t_g
= 0.36$  ms. At the strength of radial confinement $\omega_{\bot}
=10^3$ rad/s we have $\gamma = 1$. For $N=4$ the soliton contains
$\approx 720$ atoms. Similar estimates for $^7$Li condensate with
$a_s=-1.6$ nm gives $l_g \approx 7\mu$m, $t_g = 0.84$ ms,
$\omega_{\bot} =10^4$ rad/s, the soliton contains $\approx 1400$
atoms.

\section{Variational approximation}

For arbitrary forms of the reflecting potential $V(x)$, the
governing Eq. (\ref{gpe2}) cannot be analytically investigated.
One of the efficient approaches to the problem in such cases is
the variational approximation (VA), first developed for pulse
propagation in optical fibers \cite{anderson1983}, and later
applied to many other areas of nonlinear physics
\cite{malomed2002}.

Below we develop the VA for the governing equation using the
second choice (b) for the potential Eq. (\ref{mirror}). It is well
known from quantum mechanics textbooks that the wave packet
falling on the delta potential barrier is always partially
transmitted. However, by increasing the strength of the barrier
($V_0$) the transmission coefficient can be reduced to negligible
level. This allows us to consider the norm of the wave packet
above the mirror as a conserved quantity and develop the VA using
an appropriate ansatz for the pulse shape.

Eq. (\ref{gpe2}) can be generated from the following Lagrangian
density
\begin{equation}\label{lagr1}
{\cal L}=\frac{i}{2} (\psi
\psi^*_t-\psi^*\psi_t)+\frac{1}{2}|\psi_x|^2+\alpha \, x
\,|\psi|^2 - V(x)|\psi|^2-\frac{\gamma}{2} |\psi|^4.
\end{equation}
An important step in the development of VA is the proper choice of
the trail function. We shall consider the following hyperbolic
secant ansatz
\begin{equation}\label{ansatz}
\psi(x,t)=A \ {\rm sech}\left(\frac{x-\zeta}{a} \right)
e^{ib(x-\zeta)^2+i\xi(x-\zeta)+i\varphi},
\end{equation}
where $A(t), a(t), \zeta(t), \xi(t), b(t), \varphi(t)$ are
variational parameters representing the amplitude, width, center
of mass position, velocity, chirp parameter and phase of the wave
packet, respectively. This choice is motivated by the fact that
when the wave packet is sufficiently far from the reflecting
potential $V(x)$ (and therefore its effect can be neglected), Eq.
(\ref{gpe2}) has the exact accelerated soliton solution of the
hyperbolic secant form \cite{chen1976}.

\begin{figure}[htb]
\centerline{
\includegraphics[width=5.5cm,height=4cm,clip]{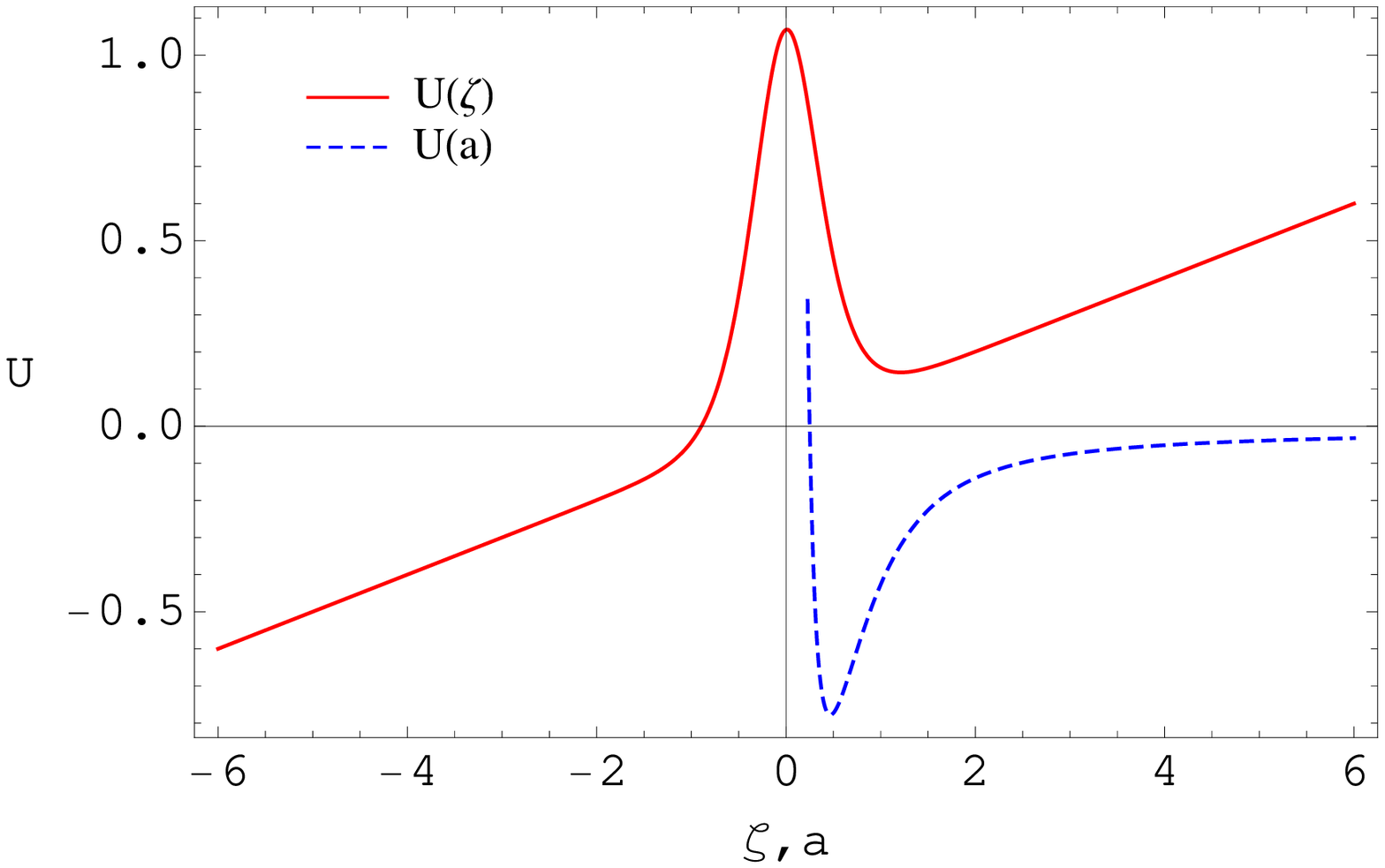} \quad
\includegraphics[width=5.5cm,height=4cm,clip]{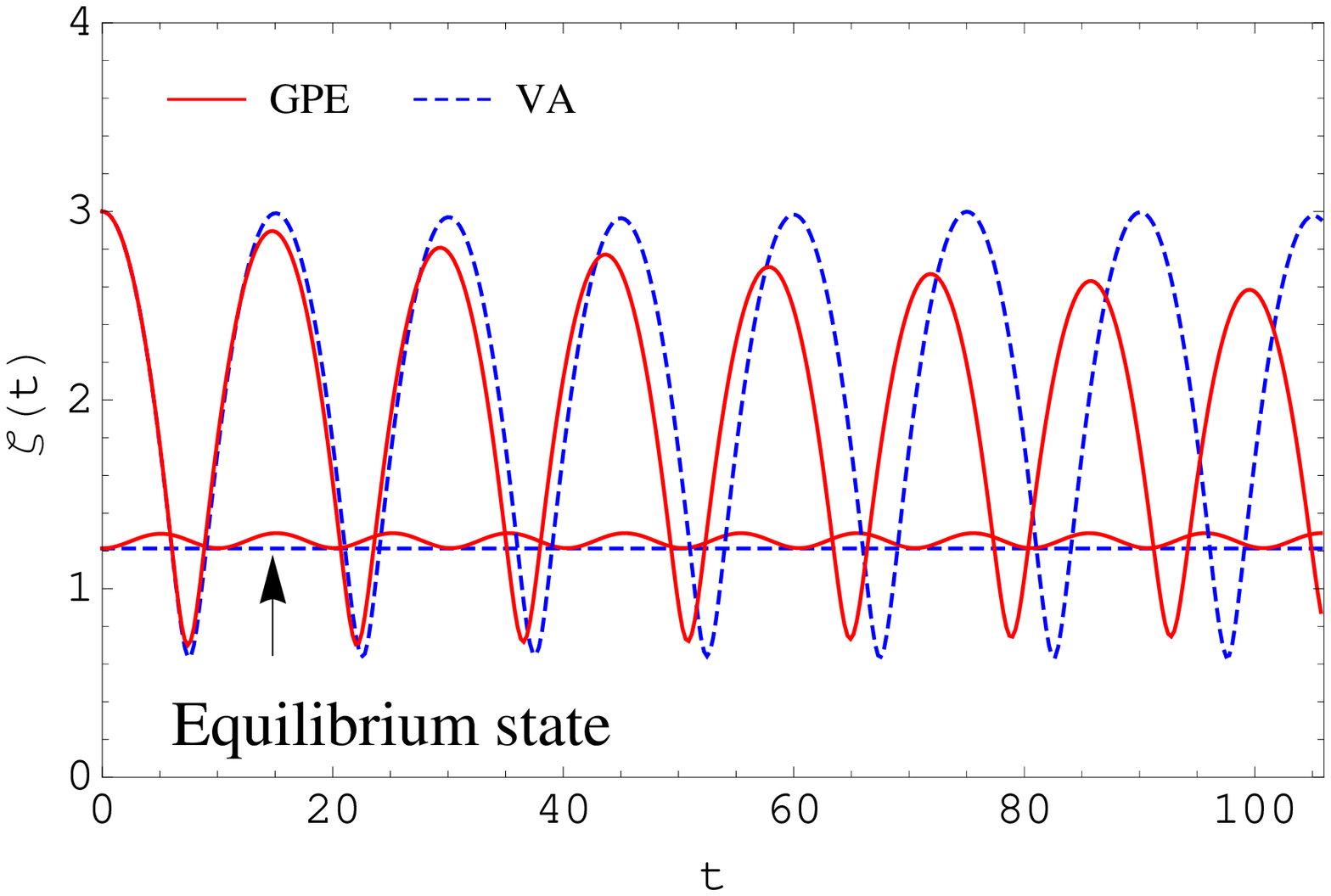} \quad
\includegraphics[width=5.5cm,height=4cm,clip]{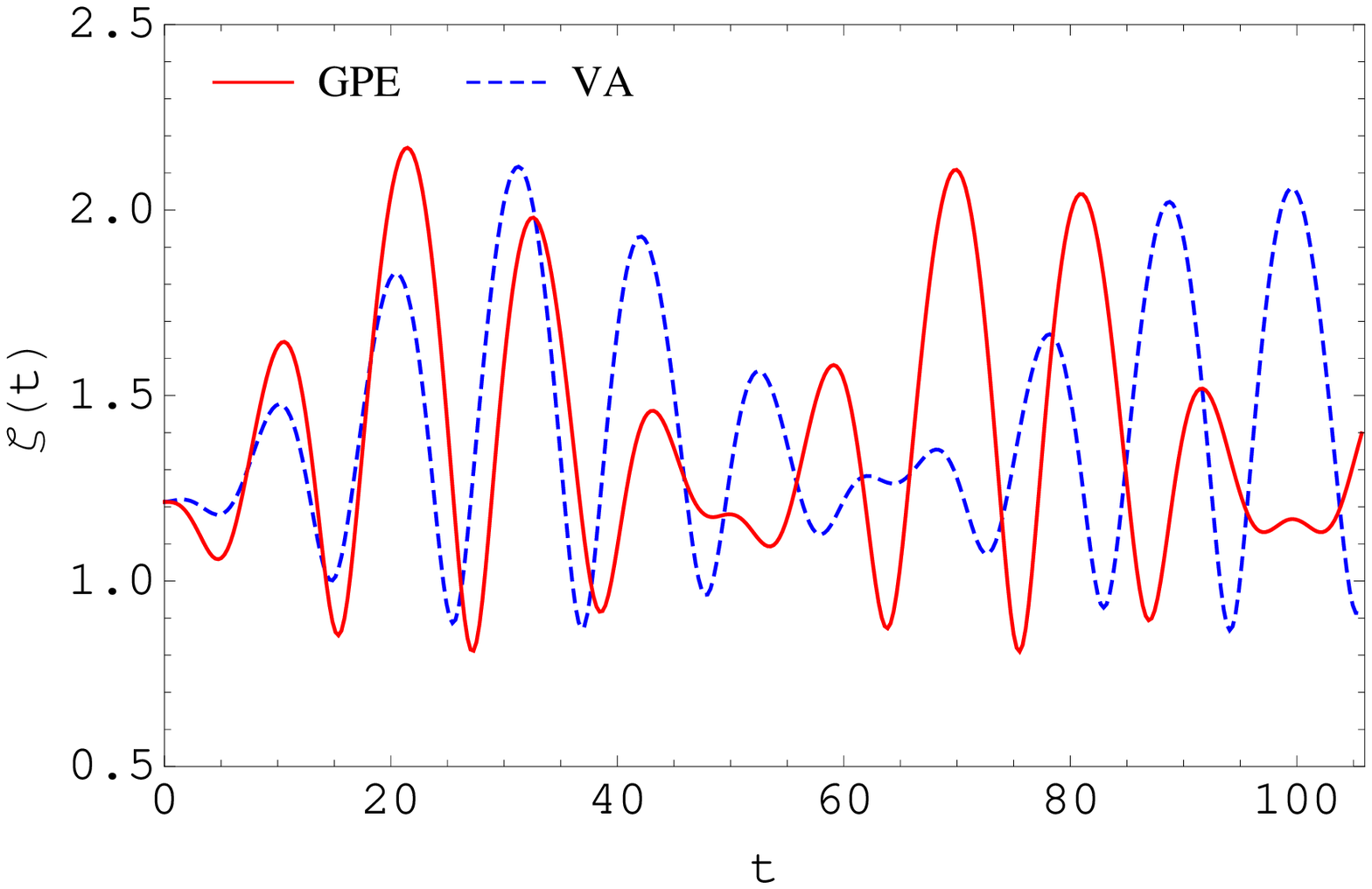}}
\caption{(Color online) Left panel: Anharmonic potentials for the
center of mass $U(\zeta)$ and width $U(a)$ of the soliton,
according to Eqs. (\ref{U1})-(\ref{U2}). For the set of parameters
$N=4$, $\gamma=1$, $\alpha = 0.1$, and $V_0=1$ the fixed point is
found to be $\zeta_0 = 1.213$, $a_0 = 0.468$. Middle panel:
Comparison of the center of mass position as a function of time,
obtained from solving the VA Eq. (\ref{att}) and numerical
simulation of the governing Eq. (\ref{gpe2}) for the reflecting
surface of the delta function type $V(x) = \alpha \, x$. The lower
pair of curves correspond to the fixed point initial conditions,
while the upper pair of curves correspond to dropping the wave
packet from height $x_0 =3$ above the mirror. Right panel:
Nonlinear resonance in the center of mass dynamics when the
coefficient of gravity is varied in time with a resonance
frequency $\alpha(t) = 0.1 [1 + 0.3 \sin(\omega_0 t)]$. Stationary
state of the soliton with parameters predicted by VA is used as
initial condition. Discrepancy (phase shift) between the GPE and
VA is associated with asymmetric deformation of the wave packet
when reflecting from the mirror.} \label{fig3}
\end{figure}

Substituting the ansatz (\ref{ansatz}) into Eq. (\ref{lagr1}) and
integrating over the space variable we get the averaged Lagrangian
$L = \int^{\infty}_{-\infty} {\cal L} dx$
\begin{equation}\label{averlagr}
L=N\left[\frac{\pi^2}{12}\,a^2b_t+\frac{\pi^2}{6}a^2b^2-
\frac{1}{2}\,\zeta_t^2-\alpha \,
\zeta+\varphi_t+\frac{1}{6a^2}+\frac{V_0}{2a}\, {\rm sech}^2\left(
\frac{\zeta}{a}\right)-\frac{\gamma N}{6a}\right],
\end{equation}
where we have taken into account that the velocity is equal to the
time derivative of the center mass position $\xi=\zeta_t$ and
$A^2=N/(2a)$, with the norm of the wave packet
$N=\int^{\infty}_{-\infty} |\psi(x)|^2 dx$ being the conserved
quantity. Now the usual procedure of the VA, applied to Eq.
(\ref{averlagr}) leads to the following set of equations for the
width and center of mass position of the wave packet
\begin{eqnarray}
a_{tt} &=& \frac{4}{\pi^2a^3}+\frac{6V_0}{\pi^2a^2}\,{\rm
sech}^2\left(\frac{\zeta}{a}\right)\,\left[1-\frac{2\zeta}{a}
\tanh
\left(\frac{\zeta}{a}\right)\right]-\frac{2 \gamma N}{\pi^2a^2}, \label{att} \\
\zeta_{tt} &=& -\alpha + \frac{V_0}{a^2}\,{\rm
sech}^2\left(\frac{\zeta}{a}\right)\,{\rm
tanh}\left(\frac{\zeta}{a}\right). \label{ztt}
\end{eqnarray}
The coupled system of equations (\ref{att})-(\ref{ztt}) represents
the main result of this paper. Its fixed points provides the
stationary width of the soliton ($a_0$) and its distance from the
mirror ($\zeta_0$), where the actions of the gravity and repulsive
potential $V(\zeta)$ cancel each other. As a result of this
balance, the soliton placed at a fixed point remains at rest
(levitates) above the mirror. Small amplitude dynamics of the
soliton's width and center of mass position near the stationary
state can be described as motion of a unit mass particle in the
anharmonic potentials $U_1(a)$ and $U_2(\zeta)$, respectively,
\begin{eqnarray}
a_{tt} &=& -\frac{\partial U_1}{\partial a}, \qquad U_1(a)
=\frac{2}{\pi^2 a^2} - \frac{2 \gamma N}{\pi^2 a} - \frac{6
V_0}{\pi^2 a} \ {\rm sech}^2 \left(\frac{\zeta_0}{a}  \right), \label{U1}\\
\zeta_{tt} &=& -\frac{\partial U_2}{\partial \zeta}, \qquad
U_2(\zeta) = \alpha \,\zeta + \frac{V_0}{2 a_0} \ {\rm
sech}^2\left(\frac{\zeta}{a_0} \right). \label{U2}
\end{eqnarray}
In Fig. \ref{fig3} the shapes of the potentials in Eqs.
(\ref{U1})-(\ref{U2}) and examples of soliton bouncing dynamics
over the reflecting surface, modelled by a delta function, are
illustrated. As expected, when the soliton is positioned at a
fixed point ($\zeta_0$, $a_0$ ), it stays motionless (lower pair
of curves in the middle panel). Small amplitude oscillations in
PDE data is due to the fact that the VA gives approximate values
for the fixed point. When the soliton is dropped towards the
mirror from a height $x_0 = 3$, it performs bouncing motion. Slow
decay of the amplitude of oscillations and increase of its
bouncing frequency are due to partial escape of the wave packet
via tunnel effect (upper pair of curves in the middle panel).

The frequency of small amplitude oscillations of the soliton's
motion can be estimated from VA by linearizing the Eqs.
(\ref{att})-(\ref{ztt}) near the fixed point ($\zeta_0, \, a_0$).
\begin{equation}\label{freq}
\omega_0 = \left(V_0/a_0^3\right)^{1/2}\,{\rm sech}^2
(\zeta_0/a_0) \left[2\,{\rm sinh}^2(\zeta_0/a_0) -1\right]^{1/2}.
\end{equation}
The corresponding period for $V_0 = 1$, $\zeta_0 = 1.213$ and $a_0
= 0.468$ is $T_0 = 2 \pi/\omega_0 \simeq 9.7$. This is in quite
good agreement with numerical simulations of the GPE (\ref{gpe2}),
shown in the middle panel of Fig. \ref{fig3}. An expression
similar to Eq. (\ref{freq}) can be derived for the frequency of
the soliton's width.
\begin{equation}\label{freq1}
\Omega_0 = (2/\pi a_0^2)\left[3 - \gamma N a_0 + (3V_0/a_0)\, {\rm
sech}^2 (\zeta_0/a_0) \left(a_0^2+2\zeta_0^2 -4a_0\zeta_0 {\rm
tanh}(\zeta_0/a_0)-3\zeta_0^2 {\rm
sech}^2(\zeta_0/a_0)\right)\right]^{1/2}.
\end{equation}
Numerical estimate for the fixed point ($\zeta_0, \, a_0$), and
$N=4$, $\gamma=1$, $V_0 = 1$ is $T_0 = 2\pi/\Omega_0 = 1.94$,
which is also in good agreement with the results of GPE.

\section{Fermi type acceleration of a matter wave soliton}

The capability of the matter wave soliton to perform bouncing
motion above the atomic mirror, preserving its integrity, suggests
to consider the Fermi type acceleration (FA) in this system. FA is
the energy gain by a particle exposed to periodic or random
driving forces. It was proposed by Enrico Fermi \cite{fermi1949}
to explain why cosmic rays have so high energy. For the mechanical
analogue, the possibility of unbounded growth of energy by an
elastic ball bouncing vertically on a single periodically
oscillating plate, under the effect of gravity, was rigorously
proven in Ref. \cite{pustilnikov1995}. Most studies of FA of
matter waves are concerned with dynamical localization and chaotic
behavior. In our model localization of the matter wave naturally
arises from the nonlinearity of the condensate, and the parameter
space does not contain chaotic regions.

\begin{figure}[htb]
\centerline{
\includegraphics[width=5.5cm,height=4cm,clip]{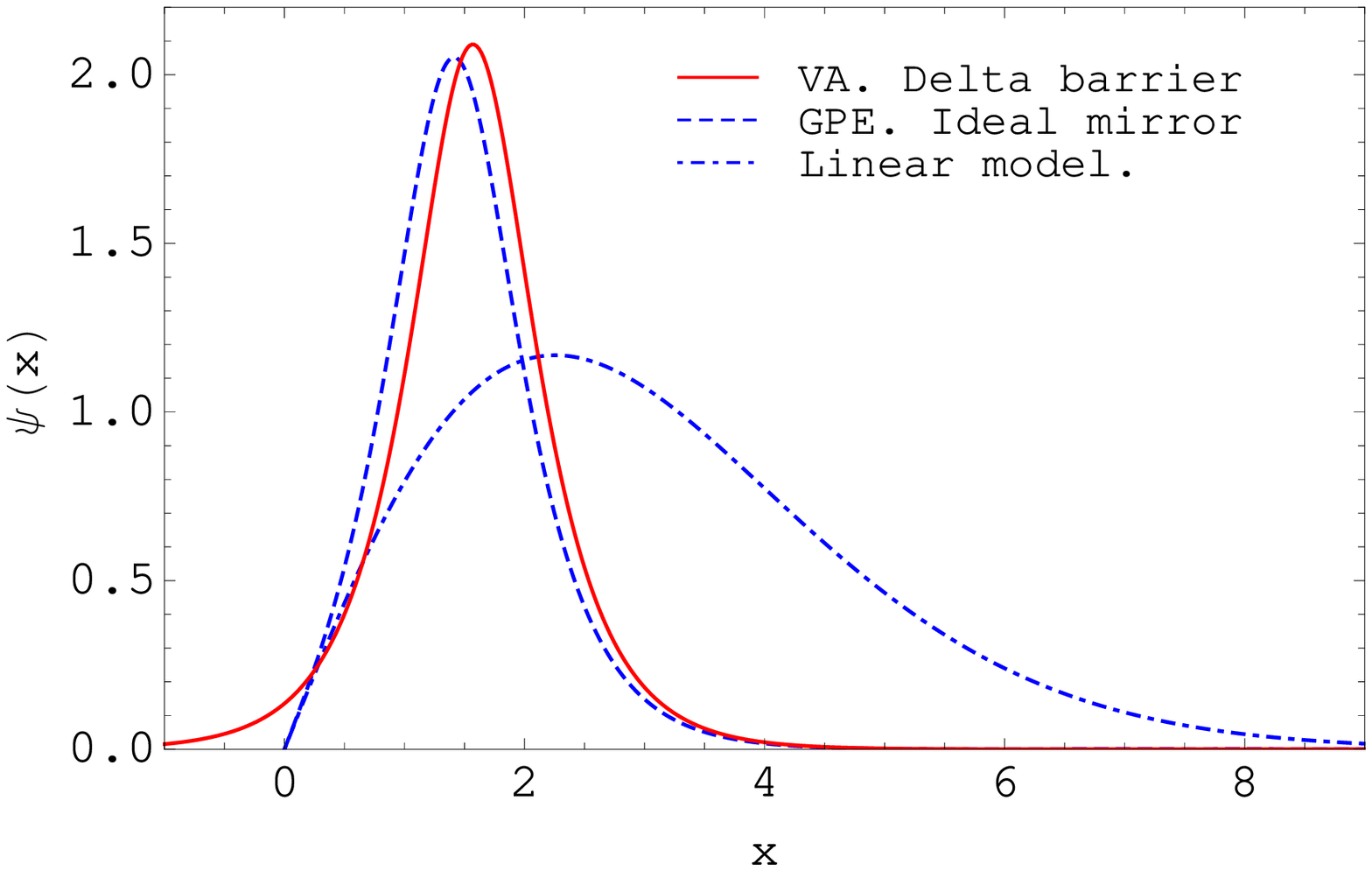} \quad
\includegraphics[width=5.5cm,height=4cm,clip]{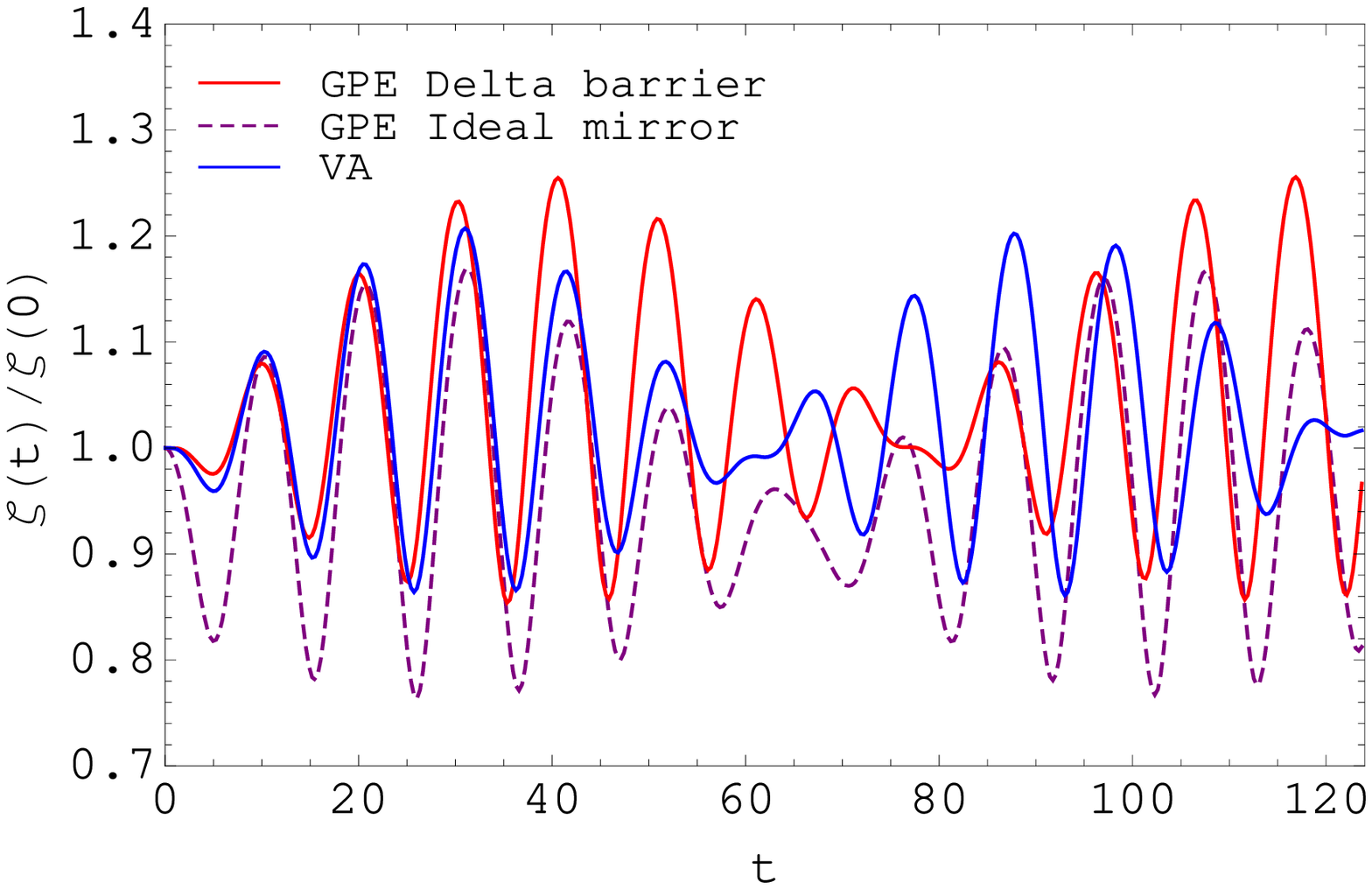} \quad
\includegraphics[width=5.5cm,height=4cm,clip]{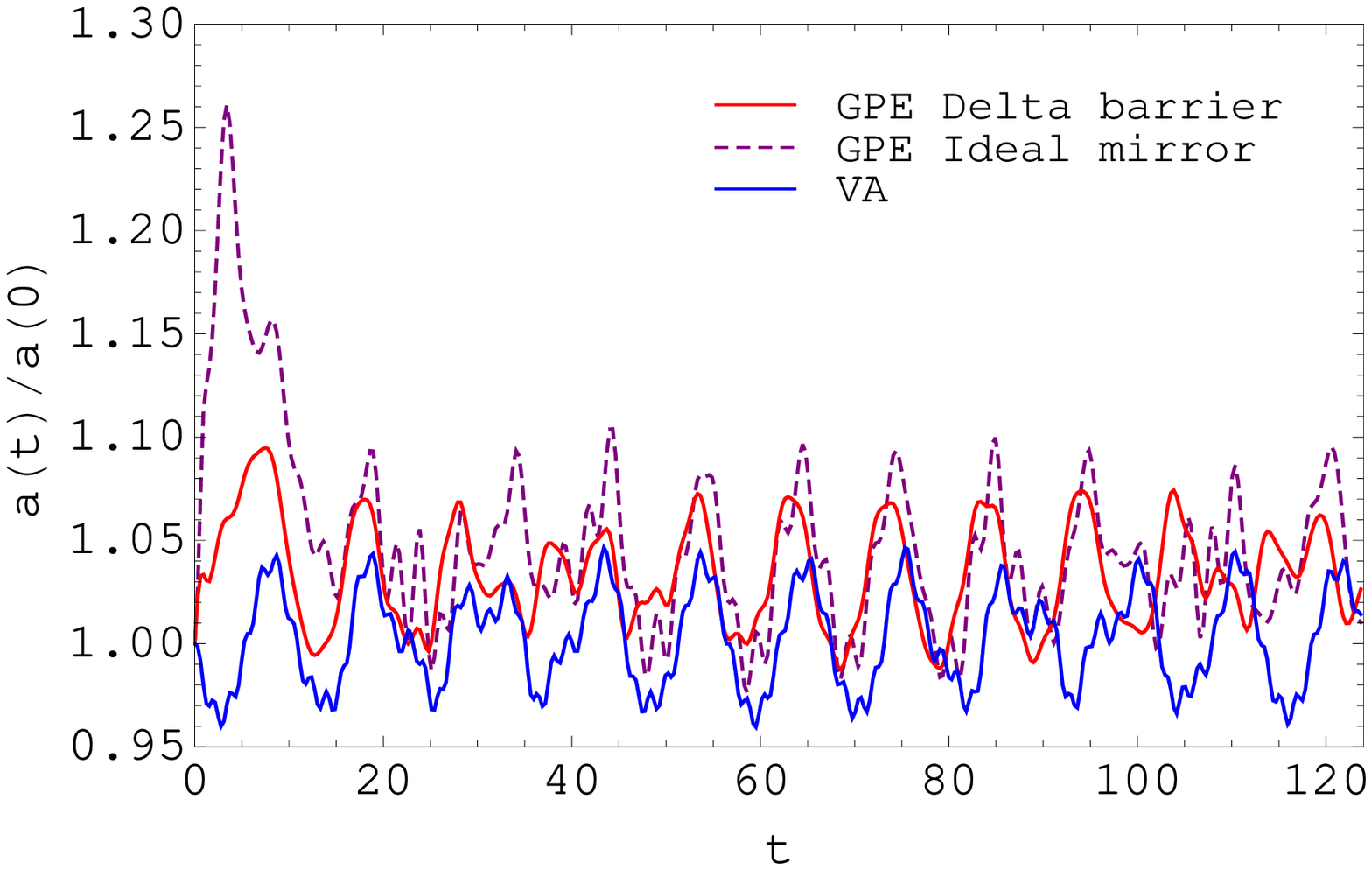}}
\caption{(Color online) Left panel: Transformation of the ground
state wave function of the linear problem (blue dot-dashed line)
into solution of the nonlinear problem (blue dashed line) by
slowly raising the coefficient of nonlinearity $\gamma$ in Eq.
(\ref{gpe2}) from zero to one. In the prediction of the VA Eqs.
(\ref{att})-(\ref{ztt}) for delta barrier potential (red solid
line) the wave packet slightly penetrates into the region $x<0$
due to the wave tunneling effect. Middle panel: Nonlinear
resonance in the center of mass dynamics of the soliton, when the
coefficient of nonlinearity is periodically varied in time $\gamma
= 1 + \epsilon \sin(\omega_0 t)$. Right panel: Dynamics of the
width has not resonant character due to the difference in
frequencies $\Omega_0$ and $\omega_0$, estimated from Eqs.
(\ref{freq})-(\ref{freq1}). Parameter values $N=4$, $\alpha =
0.1$, $V_0 = 5$, $\epsilon=0.05$, $\omega_0 = 0.66$, $\Omega_0 =
1.7$. } \label{fig4}
\end{figure}

Although the matter wave soliton does not have all necessary
properties to demonstrate true FA (due to non elastic collision
with the mirror, leakage of energy via tunnel effect, etc.),
nevertheless some features of FA can be observed, as we have
revealed in numerical experiments. At first we need to prepare the
initial stationary state of the matter wave packet levitating
above the atomic mirror. The prediction of VA for parameters of
the soliton and stationary state distance above the mirror (where
the forces of gravity and repulsion of the mirror balance out) is
approximate, as we have seen in the previous section. The
inaccuracy leads to small amplitude oscillations of the soliton
near the equilibrium state in the GPE simulations (see middle
panel in Fig. \ref{fig3}). In order to create a truly stationary
initial state of the soliton above the reflecting surface we
consider the first choice (a) for $V(x)$ in Eq. (\ref{mirror}).
For this ideal mirror potential, the Eq. (\ref{gpe2}) in the
linear limit ($\gamma=0$), with boundary condition $\psi(0,t) =
0$, has analytic stationary solutions in terms of the Airy
functions \cite{vallee2004},
\begin{equation}\label{airy}
\psi_n(x)={\cal N} \, {\rm Ai} [(2\alpha)^{1/3} \, (x + x_n)],
\end{equation}
where ${\cal N}$ is some normalization constant. Below we shall be
concerned with the ground state ($n=0$) of the wave packet in the
gravitational cavity. The first root, given by $ {\rm
Ai}[(2\alpha)^{1/3} x] = 0$ for $\alpha=0.1$, is found to be equal
to $x_0 = -3.998$. The corresponding normalization factor is
\begin{equation}\label{norm}
{\cal N}= \left( \int \limits^{\infty}_{0} {\rm Ai}^2[\alpha^{1/3}
(x+x_0)] dx \right)^{-1/2}= 1.09.
\end{equation}
To produce the initial state for numerical simulations of the
Fermi acceleration we insert the ground state wave function
(\ref{airy}) with appropriate norm into the GPE (\ref{gpe2}) with
$\gamma = 0$ and slowly raise it to final value $\gamma = 1$
according to the law $\gamma (t) = {\rm tanh} (5 t/t_0)$ with $t_0
\sim 1000$. The obtained nonlinear waveform is shown in the left
panel of Fig. \ref{fig4}.
Also in this figure we illustrate the resonant oscillations of the
soliton's center of mass when the coefficient of nonlinearity (via
atomic scattering length) is periodically changed in time. It is
evident that nonlinear resonance takes place at the frequency of
small amplitude oscillations $\omega_0$ estimated from the VA Eq.
(\ref{freq}). Similar behavior was observed when the slope of the
linear potential (strength of gravity) is changed with appropriate
frequency (see the right panel of Fig. \ref{fig3}). Since the
resonant frequencies are different for the center of mass
($\omega_0$) and width ($\Omega_0$) of the soliton, periodic
modulation of the parameter $\alpha$ or $\gamma$ with frequency
$\omega_0$ does not induce resonant oscillations of the width, and
vice versa. Characteristic feature inherent to both cases is that,
oscillations show notable phase shift as compared to predictions
of VA, which can be explained by asymmetric deformation of the
soliton at the impact with the reflecting surface. In the VA we
deal with the dynamics of a unit mass particle in the anharmonic
potential. Nevertheless the VA provides qualitatively correct
description of the system.

The focusing nonlinearity, inherent to BEC with negative s-wave
scattering length, provides the wavepacket's robustness against
dispersive spreading and different kinds of perturbations. Due to
this property matter wave solitons keep their integrity after
reflection from the atomic mirror. Below we consider the
possibility of Fermi type of acceleration in the system. In
numerical simulations we take the stationary state of the wave
packet, predicted by VA as initial condition for Eq. (\ref{gpe2})
and periodically change the vertical position of the reflecting
surface or the slope of the linear potential.

\begin{figure}[htb]
\centerline{
\includegraphics[width=8cm,height=6cm,clip]{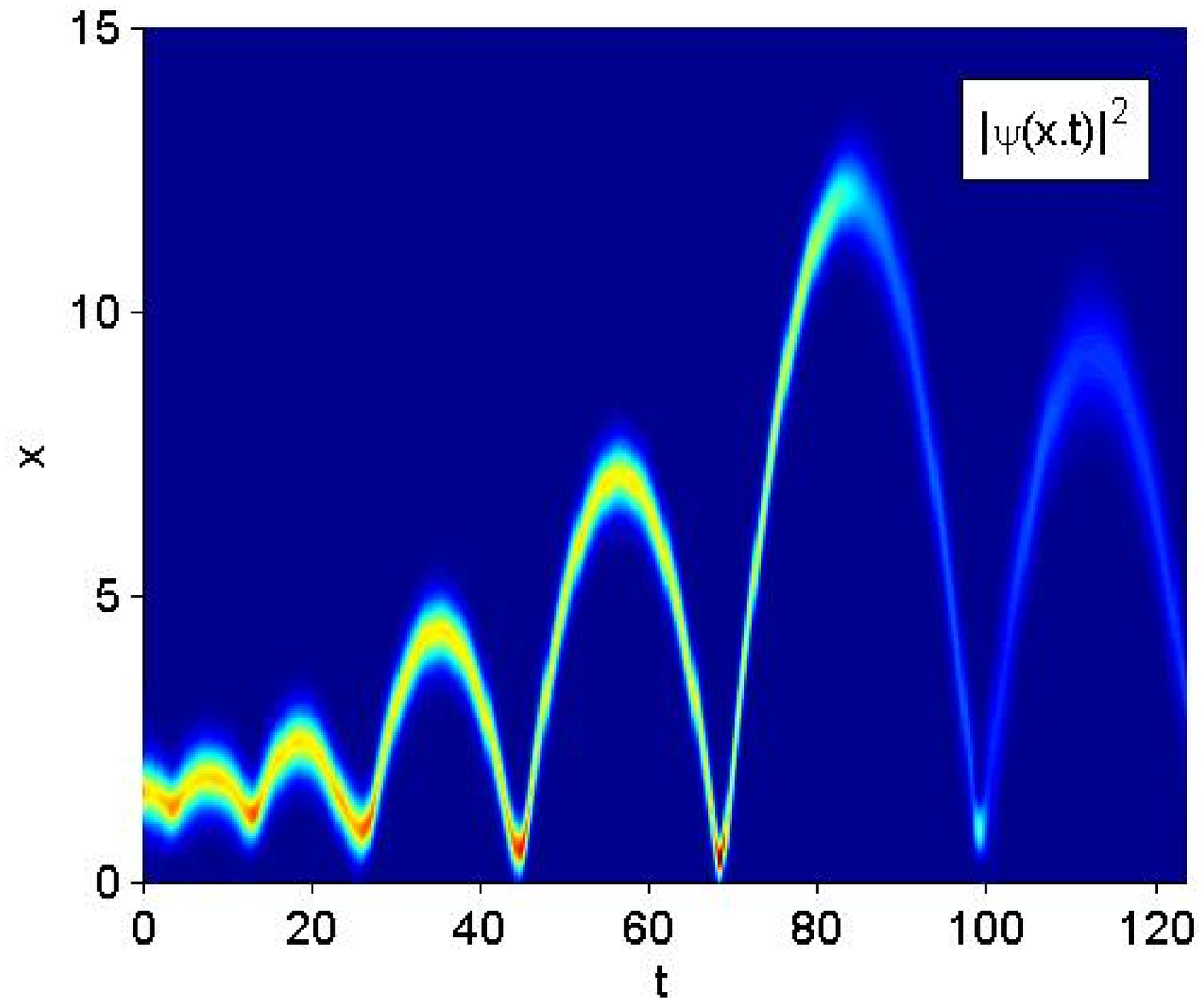} \qquad
\includegraphics[width=8cm,height=6cm,clip]{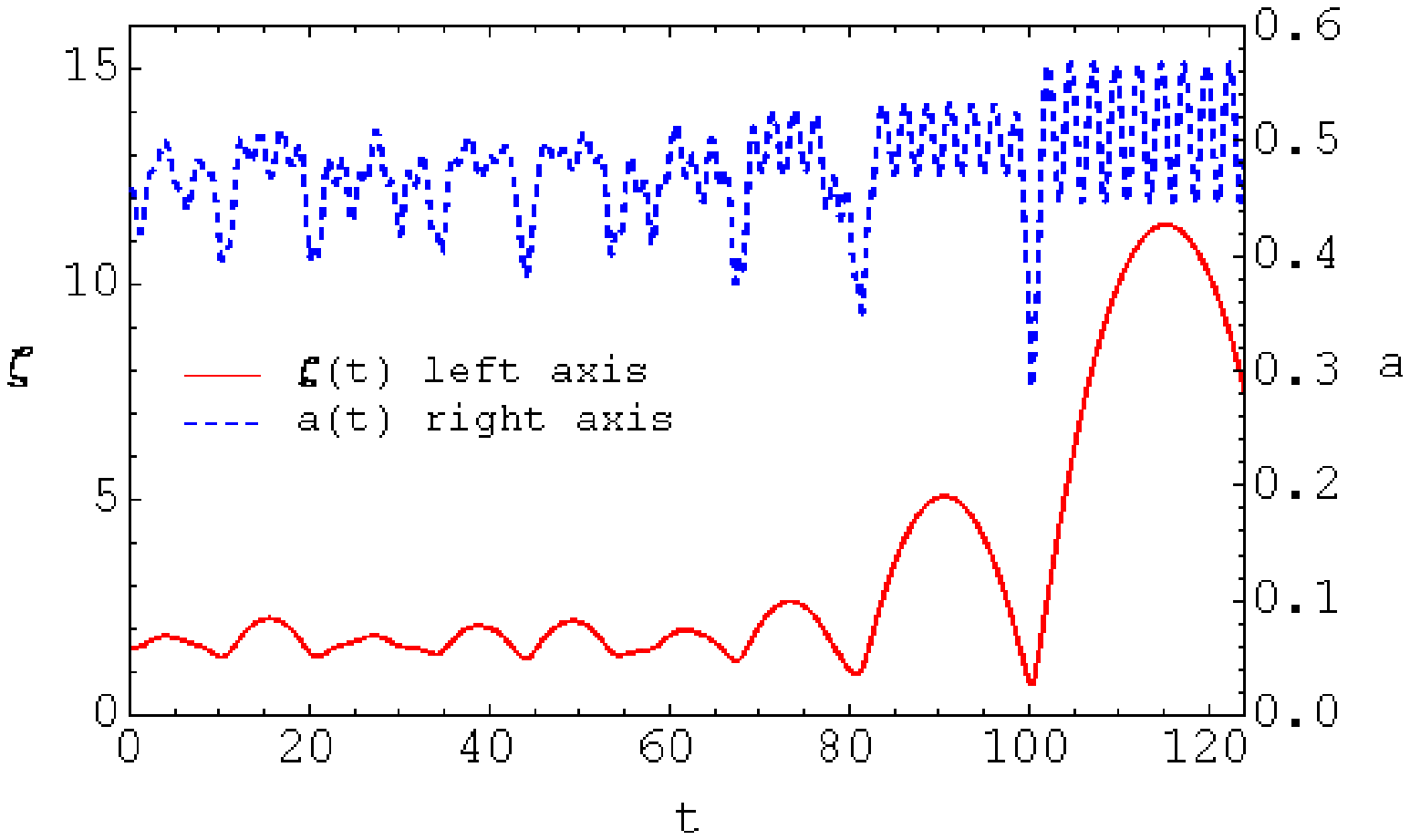}}
\caption{(Color online) Left panel: Soliton continuously increases
its kinetic energy and farther departs from the stationary point
$x_0=1.57$, when the vertical position of the delta function
mirror with strength $V_0 = 5$, initially positioned at $x=0$, is
periodically changed at a parametric resonance frequency
$f(t)=\varepsilon \, \sin(\Omega t)$, with $\varepsilon = 0.25$,
$\Omega = 2 \, \omega_0$, $\omega_0 = 0.66$, according to
numerical simulations of the GPE (\ref{gpe2}). As the amplitude of
oscillations increases, the de-tuning from the resonance occurs
and energy gain reverses. Right panel: Corresponding prediction of
the VA for the soliton's center of mass and width. A qualitative
agreement with the results of the GPE is observed. } \label{fig5}
\end{figure}

The figure \ref{fig5} illustrates the progressive gain of energy
by the soliton when the position of the reflecting delta potential
is periodically varied in time at a parametric resonance
frequency. As the amplitude of oscillation above the mirror
increases, de-tuning from the resonance occurs and further gain of
energy stops. A proper synchronization would allow more increase
of the kinetic energy of the soliton.  Also there is a
contribution of tunnel loss of the wave packet through the
reflecting delta potential barrier.

The corresponding predictions of the VA for the center of mass
position $\zeta$ and width $a$ are also shown on the right panel
of Fig. \ref{fig5}. Note that the space coordinate in GPE and VA
equations are designated by $x$ and $\zeta$ respectively.
Variation of the vertical position of the delta function mirror
$V(x) = V_0 \, \delta(x + f(t))$, where $f(t) = \varepsilon
\sin(\Omega t)$ is a periodic function with amplitude
$\varepsilon$ and frequency $\Omega$, leads to the VA equations,
similar to Eqs. (\ref{att}) - (\ref{ztt}), but with replaced space
variable on the right hand side $\zeta \rightarrow \zeta + f(t)$.
The frequency of small amplitude oscillations of the width,
measured at upper turning point is $T_0 \simeq 2.1$, which is
close to the estimation from Eq. (\ref{freq1}).

\section{Conclusions}

The model of a ``quantum bouncer" has been extended to a nonlinear
domain of Bose-Einstein condensates. The analytical description is
based on the variational approach. It has been revealed that a
matter wave soliton bouncing above the reflecting surface (or
atomic mirror) better preserves its integrity compared to a linear
wave packet due to the focusing effect of the nonlinearity. This
feature of the bright matter wave soliton allows to develop a
variational approach, using appropriate trial function, which
provides a qualitatively correct description of its dynamics. A
particle like behavior of the matter wave soliton bouncing above
the atomic mirror suggested to consider the Fermi type
acceleration in the system. In numerical experiments we observed
the progressive energy gain by the soliton when the vertical
position of the mirror is periodically varied in time. Further
development of the proposed model may include the stochastic
variation of the nonlinearity, the slope of the linear potential,
and vertical position of the reflecting surface.

\section*{Acknowledgements}
This work has been supported by the project No. CG038-2013 of the
University of Malaya. BBB is grateful to the Department of
Physics, University of Malaya, for the hospitality during his
visit.

\end{document}